\documentclass[12pt]{article}
\usepackage{amssymb,amsmath}

\begin{document}
\begin{center}
{\large {\bf Charged Particles' Tunneling from Hot-NUT-Kerr-Newman-Kasuya Spacetime}}\\
\end{center}
\vspace{2cm}
{\bf M. Hossain Ali}\\

\noindent
Department of Applied Mathematics, Rajshahi University,\\
Rajshahi-6205, Bangladesh \\
and\\
The Abdus Salam International Centre for Theoretical Physics,\\
Strada Costiera 11, 34014 Trieste, Italy\\
{\bf E-mail:} m\_hossain\_ali\_bd@yahoo.com\\

\noindent
{\it To my teacher late Prof. Mainuddin Ahmed}

\vspace{3cm}

\centerline{\bf Abstract}
\baselineskip=18pt
\bigskip

We study the Hawking radiation as charged particles' tunneling
across the horizons of the Hot-NUT-Kerr-Newman-Kasuya spacetime by
considering the spacetime background as dynamical and incorporating
the self-gravitation effect of the emitted particles when the energy
conservation, the angular momentum conservation, and the electric
charge conservation are taken into account. Our result shows that
the tunneling rate is related to the change of Bekenstein-Hawking
entropy and the radiant spectrum is not pure thermal, but is
consistent with an underlying unitary theory. The emission process
is a reversible one, and the information is preserved as a natural
result of the first law of black hole thermodynamics.
\vspace{0.5cm}

\vfill

\newpage

\section{Introduction}\label{intro}

The classical \lq \lq no hair\rq \rq  theorem stated that all
information about the collapsing body was lost from the outside
region except the three conserved quantities: the mass, the
angular momentum, and the electric charge. This loss of
information was not a serious problem in the classical theory.
Because the information could be thought of as preserved inside
the black hole but just not very accessible. However, the
situation is changed under the consideration of the quantum
effect. Black holes could shrink and eventually evaporate away
completely by emitting quantum thermal spectrum \cite{one,two}.
Since radiation with a pure thermal spectrum can carry no
information, the information carried by a physical system falling
toward black hole singularity has no way to be recovered after a
black hole has disappeared completely. This is the so-called \lq
\lq information loss paradox\rq \rq \cite{three,four}. It means
that pure quantum states (the original matter that forms the black
hole) can evolve into mixed states (the thermal spectrum at
infinity). This type of evolution violates the fundamental
principles of quantum theory which prescribe a unitary time
evolution of basis states.

The information paradox can perhaps be ascribed to the
semi-classical nature of the investigations of Hawking radiation.
However, researches in string theory indeed support the idea that
Hawking radiation can be described within a manifestly unitary
theory, but it still remains a mystery regarding the recovery of
information. Although a complete resolution of the information loss
paradox might be within a unitary theory of quantum gravity or
string/M-theory, Hawking argued that the information could come out
if the outgoing radiation were not pure thermal but had subtle
corrections \cite{four}.

Besides, there is some degree of mystery in the mechanism of black
hole radiation. In the original derivation of black hole
evaporation, Hawking described the thermal radiation as a quantum
tunneling process created by vacuum fluctuations near the event
horizon \cite{five}. But in this theory, the created mechanism of
the tunneling barrier is unclear for us. The related references do
not use the language of quantum tunneling method to discuss Hawking
radiation, and hence it is not the quantum tunneling method. In
order to derive the radiant spectrum from the black hole
horizon, one must solve the two difficulties: firstly, the formed
mechanism of the potential hill; and secondly, the elimination of
the coordinate singularity.

Recently, Kraus and Wilczek
\cite{six,seven,eight,nine,ten,eleven,twelve} did the pioneer work
for a program that implemented Hawking radiation as a tunneling
process. Parikh and Wilczek \cite{thirteen,fourteen,fifteen}
developed the program by carrying out a dynamical treatment of
black hole radiance in the static spherically symmetric black hole
geometries. More specifically, they took into account the effects
of a positive energy matter shell propagating outwards through the
horizon of the Schwarzschild and Reissner-Nordstr\"om black holes,
and incorporated the self-gravitation correction of the radiation.
In particular, they considered the energy conservation and allowed
the background geometry to fluctuate in their dynamical
description of the black hole background. In this model, they
allowed the black hole to lose mass while radiating, but
maintained a constant energy for the total system (consisting of
the black hole and the surrounding). The self-gravitation action
among the particles creates the tunneling barrier with turning
points at the location of the black hole horizon before and after
the particle with energy emission. The radiation spectrum that
they derived for the Schwarzschild and Reissner-Nordstr\"om black
holes gives a leading-order correction to the emission rate
arising from loss of mass of the black hole, which corresponds to
the energy carried by the radiated quantum. This result displays
that the radiant spectrum of the black hole is not
pure thermal under the consideration of energy
conservation and the unfixed spacetime background. This may be a
correct amendment of the Hawking radiation spectrum.

In addition to the consideration of the energy conservation and the
particle's self-gravitation, a crucial point in the analysis of
Kraus-Parikh-Wilczek (KPW) is to introduce a coordinate system that
is well-behaved at the event horizon in order to calculate the
emission probability. In this regard, the so-called \lq \lq
Painlev\'e-Gullstrand coordinates\rq \rq  rediscovered by Kraus and
Wilczek \cite{sixteen} are not only time independent and regular at
the horizon, but for which time reversal is manifestly asymptotic,
that is, the coordinates are stationary but not static. Following
this method, a lot of people
\cite{seventeen,eighteen,nineteen,twenty,twenty one,twenty
two,twenty three,twenty four,twenty five,twenty six,twenty
seven,twenty eight,twenty nine,thirty} investigated Hawking
radiation as tunneling from various spherically symmetric black
holes, and the derived results are very successful to support KPW's
picture. However, all these studies are limited to the spherically
symmetric black holes and most of them are confined only to
investigate the tunneling effect of the uncharged massless
particles. There are some attempts to extend this model to the case
of the stationary axisymmetric geometries \cite{thirty one,thirty
two,thirty three,thirty four,thirty five,thirty six,thirty
seven,thirty eight,thirty nine}. Recently, following KPW's approach,
some authors investigated the massive charged particles' tunneling
from the static spherically symmetric \cite{fourty,fourty one,fourty
two,fourty three,fourty four} as well as stationary axisymmetric
(e.g., Kerr-Newman black hole \cite{fourty five,fourty six})
geometries. They all got a satisfying result.

In this paper we apply KPW's method to a more general spacetime. We calculate the emission rate of a charged massive particle from stationary axisymmetric Kerr-Newman black hole spacetime in the de Sitter universe endowed with NUT (magnetic mass) and magnetic monopole parameters. The
metric of the spacetime can be written as
\begin{eqnarray}
\textrm{d}s^2&=&\frac{\Sigma}{\Delta_\theta}\textrm{d}\theta^2
+\frac{\Sigma}{
\Delta_r}\textrm{d}r^2-\frac{\Delta_r}{\Sigma}\left(\textrm{d}t_{HNK}-\frac{
\mathcal A}{\chi}\textrm{d}\varphi\right)^2\nonumber\\
&&+\frac{\Delta_\theta\,\textrm{sin}^2\theta}{
\Sigma}\left(a\,\textrm{d}t_{HNK}-\frac{\rho}{\chi}\textrm{d}\varphi
\right)^2\label{eq1},
\end{eqnarray}
where
\begin{eqnarray}
\Sigma&=&r^2+(n+a\,
\textrm{cos}\theta)^2,\hspace{0.5cm}\Delta_\theta=1+\frac{a^2}{\ell^2
}\,\textrm{cos}^2\theta,\hspace{0.5cm}\ell^2=\frac{3}{\Lambda},\nonumber\\
\Delta_r&=&\rho\left[1-\frac{1}{\ell^2}\,(r^2+5\,n^2)\right]-2\,(M\,r+n^2)
+Q^2+P^2,\nonumber\\
\rho&=&r^2+a^2+n^2,\hspace{0.5cm}\chi=1+\frac{a^2}{\ell^2},\hspace{0.5cm}\mathcal A=a\,\textrm{sin}^2\theta-2\,n\,\textrm{cos}\theta,\label{eq2}
\end{eqnarray}
$t_{HNK}$ being the coordinate time of the spacetime. Beside the
cosmological parameter $\Lambda$, the metric (\ref{eq1}) possesses
five parameters: $M$ the mass parameter, $a$ the angular momentum
per unit mass parameter, $n$ the NUT (magnetic mass) parameter, $Q$
the electric charge parameter, and $P$ the magnetic monopole
parameter. The metric (\ref{eq1}) solves the Einstein-Maxwell field
equations with an electromagnetic vector potential
\begin{equation}
A=-\frac{Q\,r}{\sqrt{\Sigma\,\Delta_r}}\,e^0-\frac{P\textrm{
cos}\theta}
{\sqrt{\Sigma\,\Delta_\theta}\,\textrm{sin}\theta}\,e^3\label{eq3},
\end{equation}
and an associated field strength tensor given by
\begin{eqnarray}
F&=&-\frac{1}{\Sigma^2}\left[Q\,(r^2-a^2\textrm{cos}^2\theta)
+2P\,r\,a\,\textrm{cos}\theta\right]e^0\wedge e^1\nonumber\\
&&+\frac{1}{\Sigma^2}\left[P\,(r^2-a^2\textrm{cos}^2\theta)
-2Q\,r\,a\,\textrm{cos}\theta\right]e^2\wedge e^3,\label{eq4}
\end{eqnarray}
where we have defined the vierbein field
\begin{eqnarray}
e^0&=&\sqrt{\frac{\Delta_r}{\Sigma}}\left(\textrm{d}t_{HNK}-\frac{
\mathcal A}{\chi}\textrm{d}\varphi\right),\hspace{0.5cm}e^1=\sqrt{\frac{
\Sigma}{\Delta_r}}\,\textrm{d}r,\nonumber\\
e^2&=&\sqrt{\frac{\Sigma}{\Delta_\theta}}\,\textrm{d}\theta,
\hspace{0.5cm}e^3
=\sqrt{\frac{\Delta_\theta}{\Sigma}}\,\textrm{sin}\theta\left(a\,
\textrm{d}t_{HNK}-\frac{\rho}{\chi
}\textrm{d}\varphi\right)\label{eq5}.
\end{eqnarray}

The metric (\ref{eq1}) describes the NUT-Kerr-Newman-Kasuya-de
Sitter spacetime. Since the de Sitter spacetime has been interpreted
as being hot \cite{fourty seven}, we call the spacetime a
Hot-NUT-Kerr-Newman-Kasuya (H-NUT-KN-K, for briefness) spacetime.

There is a renewed interest in the cosmological parameter as it is
found to be present in the inflationary scenario of the early
universe. In this scenario the universe undergoes a stage where it
is geometrically similar to de Sitter space \cite{fourty eight}.
Among other things inflation has led to the cold dark matter. If the
cold dark matter theory proves correct, it would shed light on the
unification of forces \cite{fourty nine,fifty}. The monopole
hypothesis was propounded by Dirac relatively long ago. The
ingenious suggestion by Dirac that magnetic monopole does exist was
neglected due to the failure to detect such object. However, in
recent years, the development of gauge theories has shed new light
on it.

The H-NUT-KN-K spacetime includes, among others, the physically
interesting black hole spacetimes as well as the NUT spacetime which
is sometimes considered as unphysical. The curious properties of the
NUT spacetime induced Misner \cite{fifty one} to consider it \lq \lq
as a counter example to almost anything\rq \rq . This spacetime
plays a significant role in exhibiting the type of effects that can
arise in strong gravitational fields.

If we set $\ell\rightarrow\infty,\,a=0,\,Q=0=P$ in Eq.(\ref{eq1}),
it then results the NUT metric which is singular along the axis of
symmetry $\theta=0$ and $\theta=\pi$. Because of the axial
singularities the metric admits different physical interpretations.
Misner \cite{fifty two} introduced a periodic time coordinate to
remove the singularity, but this makes the metric an uninteresting
particle-like solution. To avoid a periodic time coordinate, Bonnor
\cite{fifty three} removed the singularity at $\theta=0$ and related
the singularity at $\theta=\pi$ to a semiinfinite massless source of
angular momentum along the axis of symmetry. This is analogous to
representing the magnetic monopole in electromagnetic theory by
semiinfinite solenoid \cite{fifty four}. The singularity along
$z$-axis is analogous to the Dirac string.

McGuire and Ruffini \cite{fifty five} suggested that the spaces
endowed with the NUT parameter should never be directly physically
interpreted. To make a physically reasonable solution Ahmed
\cite{fifty six} used Bonnor's interpretation of the NUT parameter,
i.e., the NUT parameter $n$ is due to the strength of the physical
singularity on $\theta=\pi$, and further considered that $n=a$. That
means, the angular momentum of the mass $M$ and the angular momentum
of massless rod coalesce, and in this case, the metric (\ref{eq1})
gives a new black hole solution which poses to solve an outstanding
problem of thermodynamics and black hole physics.

In view of all the above considerations the work of this paper is
interesting. Since we are investigating charged particles' tunneling
from the charged H-NUT-KN-K spacetime, not only should the energy
conservation but also the electric charge conservation be
considered. In particular, two significant points of this paper are
as follows. The first is that we need to find the equation of motion
of a charged massive tunneling particle. We can treat the massive
charged particle as a de Broglie wave, and then its equation of
motion can be obtained by calculating the phase velocity of the de
Broglie wave corresponding to the outgoing particle. Secondly, we
should also consider the effect of the electromagnetic field outside
the H-NUT-KN-K spacetime when a charged particle tunnels out. The
Lagrangian function of the electromagnetic field corresponding to
the generalized coordinates described by $A_\mu$ is
$-\frac{1}{4}F_{\mu\nu}F^{\mu\nu}$. But these are ignorable
coordinates in dragged coordinate system. To eliminate the freedoms
corresponding to these coordinates, we modify the Lagrangian
function. Using WKB method we then derive the emission rate for a
charged massive particle with revised Lagrangian function.

We organize the paper as follows. In section \ref{sec:2} we
introduce the Painlev\'e-H-NUT-KN-K coordinate system, and obtain
the radial geodesic equation of a charged massive particle. In
section \ref{sec:3} we use KPW's tunneling framework to calculate
the emission spectrum. Finally, in section \ref{sec:4} we present
our concluding remarks. Throughout the paper, the geometrized units
($G=c=\hbar=1$) have been used.
\section{Painlev\'e-like Coordinate Transformation and the Radial Geodesics}
\label{sec:2} The null surface equation
${\textrm g}^{\mu\nu}\partial_\mu f\partial_\nu f=0$ gives
\begin{equation}
r^4+(a^2+6n^2-\ell^2)r^2+2M\ell^2r-\Xi=0\label{eq6},
\end{equation}
where
\begin{equation}
 \Xi=\left\{(a^2-n^2+Q^2 +P^2)\ell^2-5n^2(a^2+n^2)\right\}\label{eq7}.
\end{equation}
Equation (\ref{eq6}) has four real roots: one is negative root $r_-$
and three roots $r_0,\,r_H,\,r_C$ corresponding to the inner, outer
(event) and cosmological horizon of the H-NUT-KN-K spacetime
respectively, can be given by \cite{fifty seven}
\begin{eqnarray}
&&r_0=-t_1+t_2+t_3,\nonumber\\
&&r_H=t_1-t_2+t_3,\nonumber\\
&&r_C=t_1+t_2-t_3\label{eq8},
\end{eqnarray}
where
\begin{eqnarray}
&&t_1=\left[\frac{1}{6}(\ell^2-6n^2-a^2)
+\frac{1}{6}\sqrt{\{(\ell^2-6n^2-a^2)^2-12\Xi\}}
\textrm{cos}\frac{\psi}{3}\right]^{\frac{1}{2}},\nonumber\\
&&t_2=\left[\frac{1}{6}(\ell^2-6n^2-a^2)
-\frac{1}{6}\sqrt{\{(\ell^2-6n^2-a^2)^2-12\Xi\}
}\textrm{cos}\frac{\psi+\pi}{3}
\right]^{\frac{1}{2}},\nonumber\\
&&t_3=\left[\frac{1}{6}(\ell^2-6n^2-a^2)
-\frac{1}{6}\sqrt{\{(\ell^2-6n^2-a^2)^2-12\Xi\}
}\textrm{cos}\frac{\psi-\pi}{3}
\right]^{\frac{1}{2}}\label{eq9},\\
&&\textrm{cos}\psi=-\frac{(\ell^2-6n^2-a^2)\{(\ell^2-6n^2-a^2)^2
+36\Xi\}-54M^2\ell^4
}{\{(\ell^2-6n^2-a^2)^2-12\Xi\}^{3/2}}\label{eq10},
\end{eqnarray}
under the conditions
\begin{eqnarray}
\{(\ell^2-6n^2-a^2)^2-12\Xi\}^3&>&\{(\ell^2-6n^2-a^2)^3
+36\Xi(\ell^2-6n^2-a^2)\nonumber\\
&&-54M^2\ell^4\}^2,\nonumber\\
(\ell^2-6n^2-a^2)&>&0\label{eq11},
\end{eqnarray}
$\Xi$ being given by Eq.(\ref{eq7}).

To investigate the tunneling process, we should adopt the dragged
coordinate system. The line element in the dragging coordinate
system is \cite{fifty seven}
\begin{equation}
\textrm{d}s^2=\hat{\textrm g}_{00}\textrm{d}t_d^2
+\frac{\Sigma}{\Delta_r}\textrm{d}r^2
+\frac{\Sigma}{\Delta_\theta}\textrm{d}\theta^2\label{eq12},
\end{equation}
where
\begin{equation}
\hat{\textrm g}_{00}={\textrm g}_{00}
-\frac{({\textrm g}_{03})^2}{{\textrm g}_{33}}
=-\frac{\Delta_\theta\Delta_r(\rho-a\mathcal A)^2\textrm{sin}^2\theta}
{\Sigma(\Delta_\theta\rho^2\textrm{sin}^2\theta
-\Delta_r\mathcal A^2)}\label{eq13}.
\end{equation}
In fact, the line element (\ref{eq12}) represents a three-dimensional hypersurface in the four-dimensional H-NUT-KN-K spacetime. The components of the electromagnetic potential in the dragged coordinate system can be given by
\begin{equation}
A^\prime_0=A_a(\partial_{t_d})^a=-\frac{Qr}
{\Sigma}\left(1-\frac{\mathcal A}{\chi}\Omega\right)
-\frac{P\textrm{cos}\theta}{\Sigma}
\left(a-\frac{\rho}{\chi}\Omega\right), \quad
A^\prime_1=0=A^\prime_2\label{eq14},
\end{equation}
where
\begin{equation}
(\partial_{t_d})^a=(\partial_{t_{HNK}})^a+
\Omega(\partial_\varphi)^a\label{eq15},
\end{equation}
$\Omega=-{\textrm g}_{03}/{\textrm g}_{33}$ being the dragged angular
velocity. The metric (\ref{eq12}) has a coordinate singularity at
the horizon, which brings us inconvenience to investigate the
tunneling process across the horizon.

In order to eliminate the coordinate singularity from the metric
(\ref{eq12}), we perform general Painlev\'e coordinate
transformation \cite{fifty eight}
\begin{equation}
\textrm{d}t_d=\textrm{d}t+F(r,\,\theta)\textrm{d}r
+G(r,\,\theta)\textrm{d}\theta\label{eq16},
\end{equation}
which reduces the metric in the Painlev\'e-H-NUT-KN-K coordinate
system \cite{fifty seven}
\begin{eqnarray}
\textrm{d}s^2&=&\hat{{\textrm g}}_{00}\textrm{d}t^2
+\textrm{d}r^2\pm2\sqrt{\hat{\textrm g}_{00}(1-{\textrm g}_{11})}\,
\textrm{d}t\textrm{d}r+\left[\hat{\textrm g}_{00}\{G(r,\theta)\}^2
+{\textrm g}_{22}\right]\textrm{d}\theta^2\nonumber\\
&&+2\sqrt{\hat{\textrm g}_{00}(1-{\textrm g}_{11})}\,G(r,
\theta)\textrm{d}r\textrm{d}\theta+2\hat{\textrm g}_{00}G(r,
\theta)\textrm{d}t\textrm{d}\theta\label{eq17},
\end{eqnarray}
where $F(r,\,\theta)$ satisfies
\begin{equation}
{\textrm g}_{11}+\hat{{\textrm g}}_{00}\{F(r,
\theta)\}^2=1\label{eq18},
\end{equation}
and $G(r,\,\theta)$ is determined by
\begin{equation}
G(r, \theta)=\int\frac{\partial
F(r,\,\theta)}{\partial\theta}\textrm{d}r+C(\theta)\label{eq19},
\end{equation}
where $C(\theta)$ is an arbitrary analytic function of $\theta$. The
plus(minus) sign in (\ref{eq17}) denotes the spacetime line element
of the charged massive outgoing(ingoing) particles at the horizon.

According to Landau's theory of the coordinate clock synchronization
\cite{fifty nine} in a spacetime decomposed in $(3+1)$, the
coordinate time difference of two events which take place
simultaneously in different locations, is
\begin{equation}
\Delta T=-\int\frac{{\textrm g}_{0i}}{{\textrm g}_{00}}\textrm{d}x^i,\quad
(i=1,\,2,\,3)\label{eq20}.
\end{equation}
If the simultaneity of coordinate clocks can be transmitted from one
location to another and has nothing to do with the integration path,
then
\begin{equation}
\frac{\partial}{\partial
x^i}\left(-\frac{{\textrm g}_{0j}}{\hat{{\textrm g}}_{00}}\right)
=\frac{\partial}{\partial
x^j}\left(-\frac{{\textrm g}_{0i}}{\hat{{\textrm g}}_{00}}\right),\quad
(i,\,j=1,\,2,\,3)\label{eq21}.
\end{equation}
Condition (\ref{eq21}) with the metric (\ref{eq17}) gives
$\partial_\theta F(r,\,\theta)=\partial_r G(r,\,\theta)$, which is
the condition (\ref{eq19}). Thus the Painlev\'e-H-NUT-KN-K metric
(\ref{eq17}) satisfies the condition of coordinate clock
synchronization. Apart from that, the new line element has many
other attractive features: firstly, the metric is regular at the
horizons; secondly, the infinite red-shift surface and the horizons
are coincident with each other; thirdly, spacetime is stationary;
and fourthly, constant-time slices are just flat Euclidean space in
radial. All these characteristics would provide superior condition
to the quantum tunneling radiation.

The component of the electromagnetic potential in the
Painlev\'e-H-NUT-KN-K coordinate system is
\begin{equation}
A_0=-\frac{Qr} {\Sigma}\left(1-\frac{\mathcal A}{\chi}\Omega\right)
-\frac{P\textrm{cos}\theta}{\Sigma}
\left(a-\frac{\rho}{\chi}\Omega\right), \quad
A_1=0=A_2\label{eq22},
\end{equation}
which on the event horizon becomes
\begin{equation}
\left.A_0\right|_{r_H}=-V_0=-\frac{Q\,r_H}{r_H^2+a^2+n^2}, \quad
A_1=0=A_2\label{eq23}.
\end{equation}

Now let us derive the geodesics for the charged massive particles.
Since the world-line of a massive charged quanta is not light-like,
it does not follow radial light-like geodesics when it tunnels
across the horizon. For the sake of simplicity, we consider the
outgoing  massive charged particle as a massive charged shell (de
Broglie $s$-wave). According to the WKB formula, the approximative
wave function is
\begin{equation}
\phi(r,\,t)=N\textrm{exp}\left[\textrm{i}
\left(\int_{r_i-\varepsilon}^rp_r\textrm{d}r -\omega
t\right)\right]\label{eq24},
\end{equation}
where $r_i-\varepsilon$ denotes the initial location of the
particle. If
\begin{equation}
\int_{r_i-\varepsilon}^rp_r\textrm{d}r -\omega t=\phi_0\label{eq25},
\end{equation}
then, we have
\begin{equation}
\frac{\textrm{d}r}{\textrm{d}t}=\dot{r}=\frac{\omega}{k}\label{eq26},
\end{equation}
where $k$ is the de Broglie wave number. By definition, $\dot{r}$ in
(\ref{eq26}) is the phase velocity of the de Broglie wave. Unlike
the electromagnetic wave, the phase velocity $v_p$ of the de Broglie
wave is not equal to the group velocity $v_g$. They have the
following relationship
\begin{equation}
v_p=\frac{\textrm{d}r}{\textrm{d}t}=\frac{\omega}{k},\quad
v_g=\frac{\textrm{d}r_c}{\textrm{d}t}
=\frac{\textrm{d}\omega}{\textrm{d}k}=2v_p\label{eq27},
\end{equation}
where $r_c$ denotes the location of the tunneling particle. Since
tunneling across the barrier is an instantaneous process, there are
two events that take place simultaneously in different places during
the process. One is the particle tunneling into the barrier, and the
other is the particle tunneling out the barrier. In terms of
Landau's condition of coordinate clock synchronization, the
coordinate time difference of these two simultaneous events is
\begin{equation}
\textrm{d}t=-\frac{{\textrm g}_{0i}}{{\textrm g}_{00}}\textrm{d}x^i
=-\frac{{\textrm g}_{01}}{{\textrm g}_{00}}\textrm{d}r_c,\quad
(\textrm{d}\theta=0=\textrm{d}\varphi)\label{eq28}.
\end{equation}
So the group velocity is
$$
v_g=\frac{\textrm{d}r_c}{\textrm{d}t}
=-\frac{{\textrm g}_{00}}{{\textrm g}_{01}},
$$
and therefore, using (\ref{eq17}), the phase velocity (the radial
geodesics) can be expressed as
\begin{equation}
\dot{r}=v_p=\frac{1}{2}v_g =\mp\frac{\Delta_r}{2}\left[\frac{
\Delta_\theta(\rho-a\mathcal A)^2\textrm{sin}^2\theta}
{\Sigma(\Sigma-\Delta_r)(\Delta_\theta\rho^2
\textrm{sin}^2\theta-\Delta_r\mathcal A^2)}\right]
^{\frac{1}{2}}\label{eq29},
\end{equation}
where the upper sign corresponds to the geodesic of the outgoing
particle near the event horizon, and the lower sign corresponds to
that of the ingoing particle near the cosmological horizon.
Moreover, if we take into account the self-gravitation of the
tunneling particle with energy $\omega$ and electric charge $q$,
then $M$ and $Q$ should be replaced with $M\mp\omega$ and $Q\mp q$
in (\ref{eq17}) and (\ref{eq29}), respectively, with the
upper(lower) sign corresponding to outgoing(ingoing) motion of the
particle.

In the subsequent section, we shall discuss Hawking radiation from
the event and cosmological horizons, and calculate the emission rate
from each horizon by tunneling process. Sine the overall picture of
tunneling radiation for the metric is very involved, we shall
consider for simplification the outgoing radiation from the event
horizon and ignore the incoming radiation from the cosmological
horizon, when we deal with the event horizon. In the similar manner,
we shall only consider the incoming radiation from the cosmological
horizon and ignore the outgoing radiation from the event horizon for
the moment when we deal with the cosmological horizon. Of course,
this assumption is reasonable as long as the two horizons separate
away very large from each other. The radius of the cosmological
horizon is very large due to a very small cosmological constant
$\Lambda$, while the event horizon considered here is relatively
very small because the Hawking radiation can take an important
effect only for tiny black hole typical of $1\sim10\,TeV$ energy in
the brane-world scenario. Hawking radiation is a kind of quantum
effect. It can be neglected and may not be observed for an
astronomical black hole with typical star mass about $10M_\odot$.

\section{Tunneling Process of Massive Charged Particles
from H-NUT-KN-K Spacetime}\label{sec:3}

In the investigation of charged massive particles' tunneling, the
effect of the electromagnetic field outside the black hole should be
taken into consideration. So the matter-gravity system consists of
the black hole and the electromagnetic field outside the black hole.
The Lagrangian of the matter-gravity system can be written as
\begin{equation}
L=L_m+L_e\label{eq30},
\end{equation}
where $L_e=-\frac{1}{4}F_{\mu\nu}F^{\mu\nu}$ is the Lagrangian
function of the electromagnetic field corresponding to the
generalized coordinate described by $A_\mu=(A_t,\,0,\,0)$ in the
Painlev\'e-H-NUT-KN-K coordinate system. In the case of a charged
particle's tunneling out, the system transits from one state to
another. But the expression of $L_e$ tells us that
$A_\mu=(A_t,\,0,\,0)$ is an ignorable coordinate. Furthermore, in
the dragging coordinate system, the coordinate $\varphi$ does not
appear in the metric expressions (\ref{eq12}) and (\ref{eq17}). That
is, $\varphi$ is also an ignorable coordinate in the Lagrangian
function $L$. In order to eliminate these two degrees of freedom
completely, the action for the classically forbidden trajectory
should be written as
\begin{equation}
S=\int_{t_i}^{t_f}(L-p_{A_t}\dot{A}_t-
p_\varphi\dot{\varphi})\textrm{d}t\label{eq31},
\end{equation}
which is related to the tunneling rate of the emitted particle by
\begin{equation}
\Gamma\sim e^{-2\,\textrm{Im}\,S}\label{eq32}.
\end{equation}
The imaginary part of the action is
\begin{eqnarray}
\textrm{Im}\,S&=&\textrm{Im}\left\{\int_{r_i}^{r_f}
\left[p_r-\frac{\dot{A}_t}{\dot{r}}p_{A_t}
-\frac{\dot{\varphi}}{\dot{r}}p_\varphi\right]
\textrm{d}r\right\}\nonumber\\
&=&\textrm{Im}\left\{\int_{r_i}^{r_f}
\left[\int_{(0,\,0,\,0)}^{(p_r,\,p_{A_t},\,p_\varphi)}
\textrm{d}p^\prime_r-\frac{{\dot{A}_t}}{\dot{r}}
\textrm{d}p^\prime_{A_t}-\frac{\dot{\varphi}}{\dot{r}}
\textrm{d}p^\prime_\varphi\right] \textrm{d}r\right\}\label{eq33},
\end{eqnarray}
where $p_{A_t}$ and $p_\varphi$ are the canonical momenta conjugate
to $A_t$ and $\varphi$, respectively.

\subsection{Tunneling from the Event Horizon}\label{sec:3.1}
If the black hole is treated as a rotating sphere and the particle
self-gravitation is taken into account, one then finds
\begin{equation}
\dot{\varphi}=\Omega^\prime_H\label{eq34},
\end{equation}
and
\begin{equation}
J^\prime=(M-\omega^\prime)a=p^\prime_\varphi\label{eq35},
\end{equation}
where $\Omega^\prime_H$ is the dragged angular velocity of the event
horizon. The imaginary part of the action for the charged massive
particle can be written as
\begin{equation}
\textrm{Im}\,S=\textrm{Im}\left\{\int_{r_{Hi}}^{r_{Hf}}
\left[\int_{(0,\,0,\,J)}^{(p_r,\,p_{A_t},\,J-\omega a)}
\textrm{d}p^\prime_r-\frac{{\dot{A}_t}}{\dot{r}}
\textrm{d}p^\prime_{A_t}-\frac{\Omega^\prime_H}{\dot{r}}
\textrm{d}J^\prime_\varphi\right] \textrm{d}r\right\}\label{eq36},
\end{equation}
where $r_{Hi}$ and $r_{Hf}$ represent the locations of the event
horizon before and after the particle with energy $\omega$ and
charge $q$ tunnels out. We now eliminate the momentum in favor of
energy by applying Hamilton's canonical equations of motion
\begin{equation}
\dot{r}=\left.\frac{\textrm{d}H}{\textrm{d}p_r}\right|_{(r;\,
A_t,\,p_{A_t};\,\varphi,\,p_\varphi)}
=\frac{1}{\chi^2}\frac{\textrm{d}(M-\omega^\prime)}
{\textrm{d}p_r}=\frac{1}{\chi^2}\frac{\textrm{d}M^\prime}
{\textrm{d}p_r},\label{eq37}
\end{equation}

\begin{equation}
\dot{A}_t=\left.\frac{\textrm{d}H}{\textrm{d}p_{A_t}}\right|_{(A_t;\,
r,\,p_r;\,\varphi,\,p_\varphi)}
=\frac{1}{\chi}V^\prime_0\frac{\textrm{d}(Q-q^\prime)}
{\textrm{d}p_{A_t}}=\frac{1}{\chi^2}\frac{(Q-q^\prime)r_H}
{r_H^2+a^2+n^2}\frac{\textrm{d}(Q-q^\prime)}
{\textrm{d}p_{A_t}}\label{eq38},
\end{equation}
where $\frac{M}{\chi^2}$ is the total energy and
$\frac{Q}{\chi}$ is the total electric charge of the H-NUT-KN-K
spacetime, and we have treated the black hole as a charged
conducting sphere to derive (\ref{eq38}) \cite{sixty}.

It follows, similar to \cite{nine,ten}, directly that a massive
charged particle tunneling across the event horizon also sees the
effective metric (\ref{eq17}), but with the replacements
$M\rightarrow M-\omega^\prime$ and $Q\rightarrow Q-q^\prime$. We
have to perform the same substitutions in Eqs.(\ref{eq8}),
(\ref{eq23}) and (\ref{eq29}). Equation (\ref{eq29}) then gives the
desired expression of $\dot{r}$ as a function of $\omega^\prime$ and
$q^\prime$. Equation (\ref{eq36}) can now be written explicitly as
follows:
\begin{eqnarray}
\textrm{Im}\,S&=&\textrm{Im}\int_{r_{Hi}}^{r_{Hf}} \left[\int
-\frac{1}{\chi^2}\frac{2\,\sqrt{\Sigma(\Sigma
-\Delta_r^\prime)(\Delta_\theta\rho^2
\textrm{sin}^2\theta-\Delta_r^\prime\mathcal A^2)}
}{\Delta_r^\prime\,\sqrt{
\Delta_\theta(\rho-a\mathcal A)^2\textrm{sin}^2\theta}}
\right.\nonumber\\
&&\times\left.\left(\textrm{d}M^\prime-\frac{Q^\prime r_H^\prime}
{{r_H^\prime}^2+a^2+n^2}\textrm{d}Q^\prime-\Omega^\prime_H
\textrm{d}J^\prime\right)\right] \textrm{d}r\label{eq39},
\end{eqnarray}
where
\begin{eqnarray}
\Delta_r^\prime&=&(r^2+a^2+n^2)\left[1-\frac{1}
{\ell^2}\,(r^2+5\,n^2)\right]-2\,(M^\prime\,r+n^2)
+{Q^\prime}^2+P^2\nonumber\\
&=&\frac{1}{\ell^2}(r-r^\prime_-)(r-r^\prime_0)
(r-r^\prime_H)(r-r^\prime_C)\label{eq40}.
\end{eqnarray}
The above integral can be evaluated by deforming the contour around
the single pole at $r=r^\prime_H$ so as to ensure that positive
energy solution decay in time. In this way, we finish the $r$
integral and obtain
\begin{eqnarray}
\textrm{Im}\,S&=&-\frac{1}{2}\int_{(\frac{M}{\chi^2},\,\frac{Q}
{\chi})}^{(\frac{M-\omega}{\chi^2},\,\frac{Q-q} {\chi})}
\frac{1}{\chi}\frac{4\pi\ell^2({r_H^\prime}^2+a^2+n^2)}{
(r^\prime_H-r^\prime_-)(r^\prime_H-r^\prime_0)
(r^\prime_C-r^\prime_H)}\nonumber\\
&&\times\left(\textrm{d}M^\prime-\frac{Q^\prime
r_H^\prime}{{r_H^\prime}^2+a^2+n^2}
\textrm{d}Q^\prime-\Omega^\prime_H \textrm{d}J^\prime\right)
\label{eq41}.
\end{eqnarray}
Completing this integration and using the entropy expression
$S_{EH}=\pi(r_H^2+a^2+n^2)/\chi$, we obtain
\begin{equation}
\textrm{Im}\,S=-\frac{1}{2}\Delta S_{EH}\label{eq42},
\end{equation}
where $\Delta S_{EH}=S^\prime_{EH}-S_{EH}$ is the difference of
Bekenstein-Hawking entropies of the H-NUT-KN-K spacetime before and
after the emission of the particle. In fact, if one bears in mind
that
\begin{equation}
T^\prime=\frac{(r^\prime_H-r^\prime_-)(r^\prime_H-r^\prime_0)
(r^\prime_C-r^\prime_H)}{4\pi\ell^2({r_H^\prime}^2+a^2+n^2)}\label{eq43},
\end{equation}
one can get
\begin{equation}
\frac{1}{T^\prime}(\textrm{d}M^\prime-V_0^\prime
\textrm{d}Q^\prime-\Omega^\prime_H
\textrm{d}J^\prime)=\textrm{d}S^\prime\label{eq44}.
\end{equation}
That means, (\ref{eq42}) is a natural result of the first law of
black hole thermodynamics. Therefore, the emission rate of the
tunneling particle is
\begin{equation}
\Gamma\sim e^{-2\,\textrm{Im}\,S}=e^{\Delta S_{EH}}\label{eq45}.
\end{equation}
Obviously, the emission spectrum (\ref{eq45}) deviates from the
thermal spectrum.

In quantum mechanics, the tunneling rate is obtained by
\begin{equation}
\Gamma(i\rightarrow f)\sim |a_{if}|^2\cdot\alpha_n\label{eq46},
\end{equation}
where $a_{if}$ is the amplitude for the tunneling action and
$\alpha_n=n_f/n_i$ is the phase space factor with $n_i$ and $n_f$
being the number of the initial and final states, respectively.
Since $S_j\sim \textrm{ln}\,n_j$, i.e., $n_j\sim e^{S_j}$,
$(j=i,\,f)$, then
\begin{equation}
\Gamma\sim \frac{e^{S_f}}{e^{S_i}}=e^{S_f-S_i}=e^{\Delta
S}\label{eq47}.
\end{equation}
Equation (\ref{eq47}) is consistent with our result obtained by
applying the KPW's semi-classical quantum tunneling process. Hence
equation (\ref{eq45}) satisfies the underlying unitary theory in
quantum mechanics, and takes the same functional form as that of
uncharged massless particles \cite{fifty seven}.

\subsection{Tunneling at the Cosmological Horizon}\label{sec:3.2}
The particle is found tunneled into the cosmological horizon
differently from the particle's tunneling behavior of the event
horizon. When the particle with energy $\omega$ and charge $q$
tunnels into the cosmological horizon, Eqs. (\ref{eq8}),
(\ref{eq17}), (\ref{eq23}) and (\ref{eq29}) should have to modify by
replacing $M$ with $(M+\omega)$ and $Q$ with $(Q+q)$ after taking
the self-gravitation action into account. Thus, after tunneling the
particle with energy $\omega$ and charge $q$ into the cosmological
horizon, the radial geodesic takes the form
\begin{equation}
\dot{r}=\frac{\Delta^{\prime\prime}_r}{2}\left[\frac{
\Delta_\theta(\rho-a\mathcal A)^2\textrm{sin}^2\theta}
{\Sigma(\Sigma-\Delta^{\prime\prime}_r)(\Delta_\theta\rho^2
\textrm{sin}^2\theta-\Delta^{\prime\prime}_r\mathcal A^2)}\right]
^{\frac{1}{2}}\label{eq48},
\end{equation}
where
\[
\Delta_r^{\prime\prime}=(r^2+a^2+n^2)\left[1-\frac{1}
{\ell^2}(r^2+5n^2)\right]-2\left\{(M+\omega)r+n^2\right\}
+(Q+q)^2+P^2.
\]
Different from the event horizon,
($-\frac{M}{\chi^2}$,\,$-\frac{Q}{\chi}$) and
($-\frac{M+\omega}{\chi^2}$,\,$-\frac{Q+q}{\chi}$) are,
respectively, the total mass and electric charge of the H-NUT-KN-K
spacetime before and after the particle with energy $\omega$ and
charge $q$ tunnels into. We treat the spacetime as a charged
conducting sphere.

The imaginary part of the action for the charged massive particle
incoming from the cosmological horizon can be expressed as follows:
\begin{eqnarray}
\textrm{Im}\,S&=&\textrm{Im}\int_{r_{Ci}}^{r_{Cf}} \left[\int
-\frac{1}{\chi^2}\frac{2\,\sqrt{\Sigma(\Sigma
-\tilde{\Delta}_r^{\prime\prime})(\Delta_\theta\rho^2
\textrm{sin}^2\theta-\tilde{\Delta}_r^{\prime\prime}\mathcal A^2)}
}{\tilde{\Delta}_r^{\prime\prime}\,\sqrt{
\Delta_\theta(\rho-a\mathcal A)^2\textrm{sin}^2\theta}}
\right.\nonumber\\
&&\times\left.\left(\textrm{d}M^\prime-\frac{Q^\prime r_C^\prime}
{{r_C^\prime}^2+a^2+n^2}\textrm{d}Q^\prime-\Omega^\prime_C
\textrm{d}J^\prime\right)\right] \textrm{d}r\label{eq49},
\end{eqnarray}
where
\begin{eqnarray*}
\tilde{\Delta}_r^{\prime\prime}&=&(r^2+a^2+n^2)\left[1-\frac{1}
{\ell^2}\,(r^2+5\,n^2)\right]-2\,(M^\prime\,r+n^2)
+{Q^\prime}^2+P^2\\
&=&\frac{1}{\ell^2}(r-r^\prime_-)(r-r^\prime_0)
(r-r^\prime_H)(r-r^\prime_C),
\end{eqnarray*}
$r_{Ci}$ and $r_{Cf}$ are the locations of the cosmological horizon
before and after the particle of energy $\omega$ and charge $q$ is
tunneling into. There exists a single pole at the cosmological
horizon in (\ref{eq49}). Carrying out the $r$ integral, we have
\begin{eqnarray}
\textrm{Im}\,S&=&-\frac{1}{2}\int_{(-\frac{M}{\chi^2},\,-\frac{Q}
{\chi})}^{(-\frac{M+\omega}{\chi^2},\,-\frac{Q+q} {\chi})}
\frac{1}{\chi}\frac{4\pi\ell^2({r_C^\prime}^2+a^2+n^2)}{
(r^\prime_C-r^\prime_-)(r^\prime_C-r^\prime_0)
(r^\prime_C-r^\prime_H)}\nonumber\\
&&\times\left(\textrm{d}M^\prime-\frac{Q^\prime
r_C^\prime}{{r_C^\prime}^2+a^2+n^2}
\textrm{d}Q^\prime-\Omega^\prime_C \textrm{d}J^\prime\right)\nonumber\\
&=&-\frac{1}{2}\Delta S_{CH}\label{eq50},
\end{eqnarray}
where $\Delta S_{CH}$ is the change in Bekenstein-Hawking entropy
during the process of emission. The tunneling rate from the
cosmological horizon is therefore
\begin{equation}
\Gamma\sim e^{-2\,\textrm{Im}\,S}=e^{\Delta S_{CH}}\label{eq51}.
\end{equation}
It also deviates from the pure thermal spectrum and is consistent
with the underlying unitary theory, and takes the same functional
form as that of uncharged massless particles \cite{fifty seven}.

\section{Concluding Remarks}\label{sec:4}
In this paper we present our investigation of tunneling radiation
characteristics of massive charged particles from a more general
spacetime, namely, the NUT-Kerr-Newman-Kasuya-de Sitter spacetime,
which we call the Hot-NUT-Kerr-Newman-Kasuya (H-NUT-KN-K, for
briefness) spacetime, since the de Sitter spacetime has the
interpretation of being hot \cite{fourty seven}. We apply KPW's
framework
\cite{six,seven,eight,nine,ten,eleven,twelve,thirteen,fourteen,fifteen}
to calculate the emission rate at the event/cosmological horizon. We
first introduce a simple but useful Painlev\'e coordinate
\cite{fifty eight} which transforms the line element in a convenient
form with having many superior features in favor of our study.
Secondly, we treat the charged massive particle as a de Broglie
wave, and derive the equation of motion by computing the phase
velocity. Thirdly, we take into account the particle's
self-gravitation and treat the background spacetime as dynamical.
Then the energy conservation and the angular momentum conservation
as well as the electric charge conservation are enforced in a
natural way. Adapting this tunneling picture we were able to compute
the tunneling rate and the radiant spectrum for a massive charged
particle with revised Lagrangian function and WKB approximation. The
result displays that tunneling rate is related to the change of
Bekenstein-Hawking entropy and depends on the emitted particle's
energy and electric charge. Meanwhile, this implies that the
emission spectrum is not perfect thermal but is in agreement with an
underlying unitary theory.

The result obtained by us reduces to the Kerr-Newman black hole case
for $\ell\rightarrow\infty$, $P=0=n$, and gives the result of Zhang
et al. \cite{fourty five}. For $\ell\rightarrow\infty$, $a=0=n$, our
result reduces to that of the Reissner-Nordstr\"om black hole, as
was obtained in \cite{twenty nine}. Moreover, by suitably choosing
the parameters of the spacetime, the result of this paper can be
specialized for all the interesting black hole spacetimes, de Sitter
spacetimes as well as the NUT spacetime which has curious properties
\cite{fifty one}. The NUT spacetime is a generalization of the Schwarzschild spacetime and plays an important role in the conceptual development of general relativity and in the construction of brane solutions in string theory and M-theory \cite{sixty one,sixty two,sixty three}. The NUT spacetime has been of particular interest in recent years because of the role it plays in furthering our understanding of the AdS-CFT correspondence \cite{sixty four,sixty five,sixty six}. Solutions of Einstein equations with negative cosmological constant $\Lambda $ and a nonvanishing NUT charge have a boundary metric that has closed timelike curves. The behavior of quantum field theory is significantly different in such spacetimes. It is of interest to understand how AdS-CFT correspondence works in these sorts of cases \cite{sixty seven}. Our result can directly be extended to the AdS case (as was obtained in Taub-NUT-AdS spacetimes in \cite{sixty eight}) by changing the sign of the cosmological parameter $\ell^2$ to a negative one. In view of these attractive features, the study of this paper is interesting.

Our study indicates that the emission process satisfies the first
law of black hole thermodynamics, which is, in fact, a combination
of the energy conservation law:
$\textrm{d}M-V_0\textrm{d}Q-\Omega_H\textrm{d}J=dQ_h$ and the second
law of thermodynamics: $\textrm{d}S=dQ_h/T$, $Q_h$ being the heat
quantity. Indeed, the equation of energy conservation is suitable
for any process, reversible or irreversible, but
$\textrm{d}S=dQ_h/T$ is only reliable for the reversible process;
for an irreversible process, $\textrm{d}S>dQ_h/T$. The emission
process in KPW tunneling framework is thus an reversible one.  In
this picture, the background spacetime and the outside approach an
thermal equilibrium by the process of entropy flux
$\textrm{d}S=dQ_h/T$. As the H-NUT-KN-K spacetime radiates, its
entropy decreases but the total entropy of the system remains
constant, and the information is preserved. But in fact, the
existence of the negative heat capacity makes an evaporating black
hole a highly unstable system, and the thermal equilibrium between
the black hole and the outside becomes unstable, there will exist
difference in temperature. Then the process is irreversible,
$\textrm{d}S>dQ_h/T$, and the underlying unitary theory is not
satisfied. There will be information loss during the evaporation and
the KPW's tunneling framework will fail to prove the information
conservation. Further, the preceding study is still a semi-classical
analysis -- the radiation is treated as point particles. The
validity of such an approximation can only exist in the low energy
regime. To properly address the information loss problem, a better
understanding of physics at the Planck scale is a necessary
prerequisite, especially that of the last stages or the endpoint of
Hawking evaporation.

Using the interesting method of complex paths Shankanarayanan et al.
\cite{sixty nine,seventy,seventy one} investigated the Hawking
radiation by tunneling approach, considering the amplitude for pair
creation both inside and outside the horizon. In their formalism the
tunneling of particles produced just inside the horizon also
contributes to the Hawking radiation.

Akhmedov et al. \cite{seventy two,seventy three} investigated Hawking
radiation in the quasi-classical tunneling picture by the
Hamilton-Jacobi equations, using
$\Gamma\propto\textrm{exp}\{\textrm{Im}(\oint p\textrm{d}r)\}$
\cite{seventy four}, rather than
$\Gamma\propto\textrm{exp}\{2\,\textrm{Im}(\int p\textrm{d}r)\}$,
and argued that the former expression for $\Gamma$ is correct since
$\oint p\textrm{d}r$ is invariant under canonical transformation,
while $\int p\textrm{d}r$ is not. According to their argument the
temperature of the Hawking radiation should be twice as large as
originally calculated.

Wu et al. \cite{seventy five} studied the tunneling effect near a
weakly isolated horizon \cite{seventy six} by applying the null
geodesic method of KPW and the Hamilton-Jacobi method \cite{thirty
one}, both lead to the same result. However, there are subtle
differences, e.g., in KPW's method, only the canonical time
direction can define the horizon mass and lead to the first law of
black hole mechanics, while the thermal spectrum exists for any
choice of time direction in the Hamilton-Jacobi method. Berezin et al. \cite{seventy seven} used a self-consistent canonical quantization of self-gravitating spherical shell to describe Hawking radiation as tunneling. Their work is analogous to KPW but due to the fact that they took into account back reaction of the shell on the metric they did not have a singular potential at $r_g$ (it is smoothed in their case between $r_{\rm in}$ and $r_{\rm out}$) and the use of the semi-classical approximation to describe tunneling seems more justified.

Since the discovery of the first exact solution of Einstein's field
equations, the studying property of black holes is always a
highlight of gravitational physics. Providing mechanisms to fuel the
most powerful engines in the cosmos, black holes are playing a major
role in relativistic astrophysics. Indeed, the famous Hawking
radiation from the event horizon of black holes is one of the most
important achievements  of quantum field theory in curved
spacetimes. In fact, due to Hawking evaporation classical general
relativity, statistical physics, and quantum field theory are
connected in quantum black hole physics. It is generally believed
that the deep investigation of black hole physics would be helpful
to set up a satisfactory quantum theory of gravity. In view of this,
tunneling process of Hawking radiation deserves more investigations
in a wider context.

\vspace{1.0cm}

\noindent
{\large\bf Acknowledgement}\\
I am thankful to SIDA as well as the Abdus Salam International
Centre for Theoretical Physics, Trieste, Italy, where this paper was
produced during my Associateship visit.

\begin{center}
\Large {$\diamond $}
\end{center}

\end{document}